\documentclass[aps,twocolumn,floats,pre]{revtex4}
\usepackage{graphics,graphicx,epsfig}
\usepackage{amssymb}
\usepackage{epsf,epstopdf,wrapfig}

\begin{document}

\title{Efficient representation as a design principle for neural coding and computation}

\author{William Bialek,$^a$ Rob R. de Ruyter van Steveninck$^b$ and 
Naftali Tishby$^c$}

\affiliation{$^a$Joseph Henry Laboratories of Physics, Lewis--Sigler Institute for Integrative Genomics, and Princeton Center for Theoretical Physics, Princeton University, Princeton, NJ 08544 USA\\
$^b$Department of Physics and Program in Neuroscience,
Indiana University, Bloomington, IN 47405, USA\\
$^c$School of Computer Science and Engineering, and
Interdisciplinary Center for Neural Computation,
Hebrew University,
Jerusalem 91904, Israel}

\begin{abstract}
Does the brain construct an efficient representation of the sensory world?  We review progress on this question, focusing on a series of experiments in the last decade which use fly vision as a model system in which theory and experiment can confront each other.  Although the idea of efficient representation has been productive, clearly it is incomplete since it doesn't tell us which bits of sensory information are most valuable to the organism.  We suggest that an organism which maximizes the (biologically meaningful) adaptive value of its actions given fixed resources should have internal representations of the outside world that are optimal in a very specific information theoretic sense:  they maximize the information about the future of sensory inputs at a fixed value of the information about their past.  This principle contains as special cases computations which  the brain seems to carry out, and it should be possible to test this optimization directly.  We return to the fly visual system and report the results of preliminary experiments that are in encouraging agreement with theory.
\end{abstract}

\date{\today}

\maketitle

\section{Introduction}

Since Shannon's original work \cite{shannon_48} there has been the hope that information theory would provide not only a guide to the design of engineered communication systems but also a framework for understanding information processing in biological systems.  One of the most concrete implementations  of this idea is the proposal that computations in the brain serve to construct an efficient (perhaps even maximally efficient) representation of incoming sensory data \cite{attneave_54,barlow_59,barlow_61}. 
Since efficient coding schemes are matched, at least implicitly, to the distribution of input signals, this means that what the brain computes---perhaps down to the properties of individual neurons---should be predictable from the statistical structure of the sensory world.  This is a very attractive picture, and points toward general theoretical principles rather than just a set of small models for different small pieces of the brain.  More precisely, this picture suggests a research program that could lead to an experimentally testable theory. 

Our research efforts, over several years, have been influenced by these ideas of efficient representation.  On the one hand, we have found evidence for this sort of optimization in the responses of single neurons in the fly visual system, especially once we developed tools for exploring the responses to more naturalistic sensory inputs.  On the other hand, we have been concerned that simple implementations of information theoretic optimization principles must be wrong, because they implicitly attach equal value to all possible bits of information about the world.  In response to these concerns, we have been trying to develop alternative approaches, still grounded in information theory but not completely agnostic about the value of information.  Guided by our earlier results, we also want to phrase these theoretical ideas in a way that suggests new experiments.  

What we have outlined here is an ambitious program, and certainly we have not reached anything like completion.  The invitation to speak at the International Symposium on Information Theory in 2006 seemed like a good occasion for a progress report, so that is what we present here. It is much easier to convey the sense of `work in progress' when speaking than when writing, and we hope that the necessary formalities of text do not obscure the fact that we are still groping for the correct formulation of our ideas.  We also hope that, incomplete as it is, others will find the current state of our understanding useful and perhaps even provocative.

\section{Some results from the fly visual system}

The idea of efficient representation in the brain has motivated a considerable amount of work over several decades.  We begin  by reviewing
some of what has been done along these lines, focusing on one experimental testing ground, the motion sensitive neurons in the fly visual system. 

Many animals, in particular those that fly,  rely on visual motion estimation to navigate through the world. The sensory--motor system responsible for this task, loosely referred to as the optomotor control loop, has been the subject of intense investigation in the fly, both in behavioral \cite{reichardt+poggio_76} and in electrophysiological studies. In particular, Bishop and Keehn \cite{bishop+keehn_66} described wide field motion sensitive cells in the fly's lobula plate, and some neurons of this class have been directly implicated in optomotor control \cite{hausen+wehrhahn_83}.  The fly's motion sensitive visual neurons thus are critical for behavior, and one  can record the action potentials or spikes generated by individual motion sensitive cells (e.g., the cell H1, a lobula plate neuron selective for horizontal inward motion) using an extracellular tungsten microelectrode, and standard electrophysiological methods \cite{spikes}; unlike most such recordings, in the fly one can record stably and continuously for days. 

The extreme stability of the H1 recordings has made  this system an attractive testing ground for a wide variety of issues in neural coding and computaton.  In particular, for H1  it has been possible to show that:
\begin{enumerate}
\itemsep = 0mm
\item Sequences of action potentials provide large amounts of information about visual inputs, within a factor of two of the limit set by the entropy of these sequences even when we distinguish spike arrival times with millisecond resolution \cite{strong+al_98}.
\item This efficiency of coding has significant contributions from temporal patterns of spikes that provide more information than expected by adding up the information carried by individual spikes \cite{brenner+al_00b}.
\item Although many aspects of the neural response vary among individual flies, the efficiency of coding is nearly constant \cite{schneidman+al_01}.
\item Information rates and coding efficiencies are higher, and the high efficiency extends to even higher time resolution, when we deliver stimulus ensembles that more closely approximate the stimuli which flies encounter in nature \cite{lewen+al_01,nemenman+al_06}.
\item The apparent input/output relation of these neurons changes in response to changes in the input distribution.  For the simple case where we change the dynamic range of velocity signals, the input/output relation rescales so that the signal is encoded in relative units; the magnitude of the rescaling factor maximizes information transfer \cite{brenner+al_00a}.
\item In order to adjust the input/output relation reliably, the system has to collect enough samples to be sure that the input distribution has changed.  In fact the speed of adaptation is close to this theoretical limit \cite{fairhall+al_01}.
\end{enumerate}
All of these results point toward the utility of efficient representation as a hypothesis guiding the design of new experiments, and perhaps even as a real theory of the neural code.  So much for the good news.

\section{On the other hand ...}

Despite the successes of information theoretic approaches to the neural code in fly vision and in other systems, we must be honest and consider the fundamental stumbling blocks in {\em any} effort to use information theoretic ideas in the analysis of biological systems.  First, Shannon's formulation of information theory has no place for the value or meaning of the information. This is not an accident.  On the first page of his 1948 paper \cite{shannon_48}, Shannon remarked (italics in the original):
\begin{quote}
Frequently the messages have {\em meaning}; that is they refer to or are correlated according to some system with certain physical or conceptual entities.  These semantic aspects of the communication are irrelevant to the engineering problem.
\end{quote}
Yet surely organisms find some bits more valuable than others, and any theory  that renders meaning irrelevant must miss something fundamental about how organisms work. Second, it is difficult to imagine that evolution can select for abstract quantities such as the number of bits that the brain extracts from its sensory inputs.  Both of these problems point away from general mathematical structures toward biological details such as the fitness or adaptive value of particular actions, the costs of particular errors, and the resources needed to carry out specific computations.  It would be attractive to have a theoretical framework that is faithful to these biological details but nonetheless derives predictions from more general principles.  

\begin{figure}[b]
\centerline{\includegraphics[width=\columnwidth]{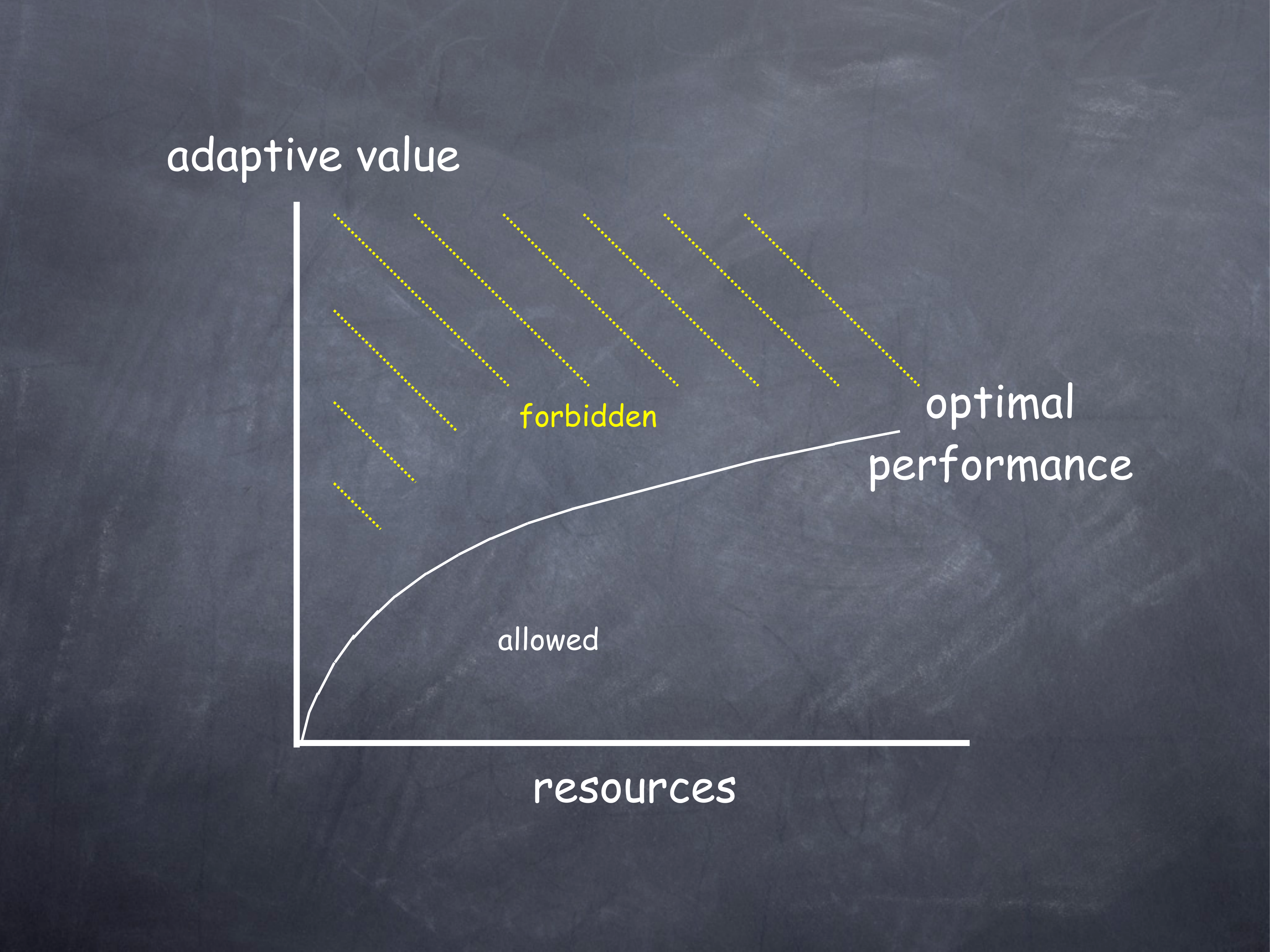}}
\caption{Optimization from a biological point of view.}
\label{bio_opt}
\end{figure}

To develop a biologically meaningful notion of optimization, we should start with the idea that there is some metric (`adaptive value' in Fig \ref{bio_opt}) for the quality or utility of the actions taken by an organism, and that there are resources that the organism needs to spend in order to take these actions and to maintain the apparatus that collects and processes the relevant sensory information.
The ultimate metric is evolutionary fitness, but in more limited contexts one can think about the frequency or value of rewards and punishments, and in experiments one can manipulate these metrics directly.  Costs often are measured in metabolic terms, but one also can  measure the volume of neural circuitry  devoted to a task.  Presumably there also are costs associated with the development of complex structures, although these are difficult to quantify.

Given precise definitions of utility and cost for different strategies (whether represented by neurons or by genomes), the biologically meaningful optimum is to maximize utility at fixed cost:  While there may be no global answer to the question of how much an organism should spend, there is a notion that it should receive the maximum return on this investment.  Thus within a given setting there is a curve that describes the maximum possible utility as a function of the cost, and this curve divides the utility/cost plane into regions that are possible and impossible for organisms to achieve, as in Fig \ref{bio_opt}.  This curve defines a notion of optimal performance that seems well grounded in the facts of life, even if we can't compute it.  The question is whether we can map this biological notion of optimization into something that has the generality and power of information theory.

\section{Costs and benefits are related to information}

\begin{figure}
\centerline{\includegraphics[width=\columnwidth]{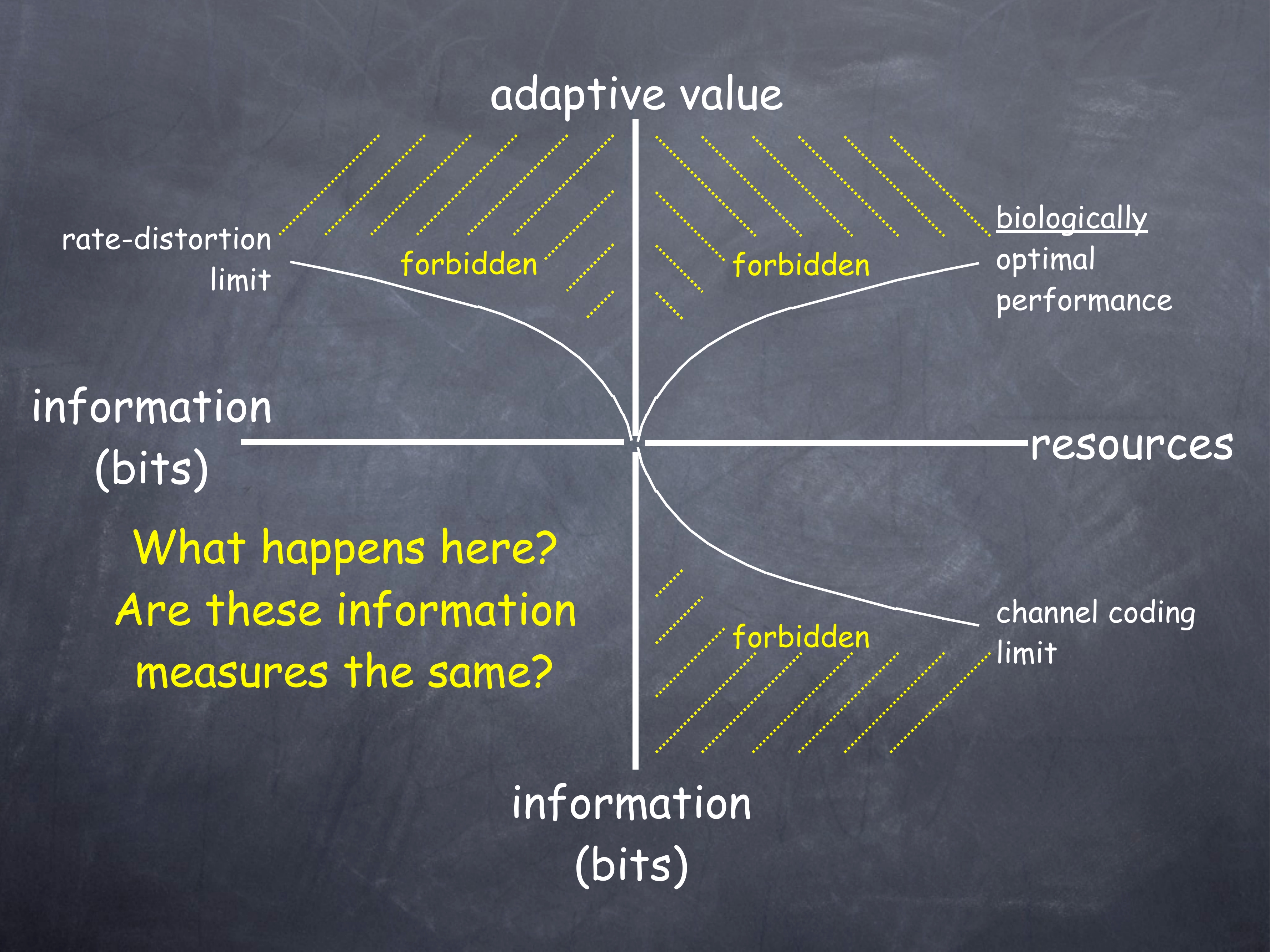}}
\caption{Biological costs and benefits are connected to bits.  The upper right quadrant redraws the biologically motivated notion of optimization from Fig \ref{bio_opt}, trading resources for adaptive value.  In the upper left, we show schematically that achieving a given quality of performance requires a minimum number of bits, in the spirit of rate--distortion theory.  In the lower right, we show that a given resource expenditure will suffice only to collect a certain maximal number of bits, in the spirit of channel coding.  Through these connections, the quantities that govern biological optimization are translated into bits.  But are these the same bits?}
\label{3quad}
\end{figure}

To begin,  we note that taking actions which achieve a criterion level of fitness requires a minimum number of bits of information, as schematized in the upper left of Fig \ref{3quad}.  Consider an experiment in which human subjects point at a target, and the reward or utility is dependent upon the positional error of the pointing.  We can think of the motor neurons, muscles and kinematics of the arm together as a communication channel that transforms some central neural representation into mechanical displacements.  If we had an accurate model of this communication channel we could calculate its {\em rate--distortion function}, which determines the minimum number of bits required in specifying the command to insure displacements of specified accuracy across a range of possible target locations.  
The  rate--distortion function  divides the utility/information plane into accessible and inaccessible regions.  

It also is true that bits are not free.  In the classical examples of communication channels, the signal--to--noise ratio ($SNR$) with which data can be transmitted is related directly to the power dissipation, and the $SNR$ in turn sets the maximum number of bits that can be transmitted in a given amount of time; this is (almost) the concept of channel capacity.  If we think about the bits that will be used to direct an action, then there are many costs---the cost of acquiring the information, of representing the information, and the more obvious physical costs of carrying out the resulting actions. Continuing with the example of motor control,  we always can assign these costs to the symbols at the entrance to the communication channel formed by the motor neurons, muscles and arm kinematics.     The channel capacity separates the information/cost plane into accessible and inaccessible regions, as in the lower right quadrant of Fig \ref{3quad}.  Ideas about metabolically efficient neural codes  \cite{laughlin+al_98,balasubramanian+al_01} can be seen as efforts to calculate this curve in specific models.  

\section{Can we close the loop?}

To complete the link between biological optimization and information theoretic ideas, we need to remember that there is a causal path from  information about the outside world to  internal representations to actions.
Thus the adaptive value of actions always depends on the state of the world {\em after} the internal representation has been formed, simply because it takes time to transform representations into actions;  the only bits that can contribute to fitness are those which have predictive power regarding the future state of the world.  In contrast, because of causality, any internal representation necessarily is built out of information about the past.  

\begin{figure}
\centerline{\includegraphics[width=\columnwidth]{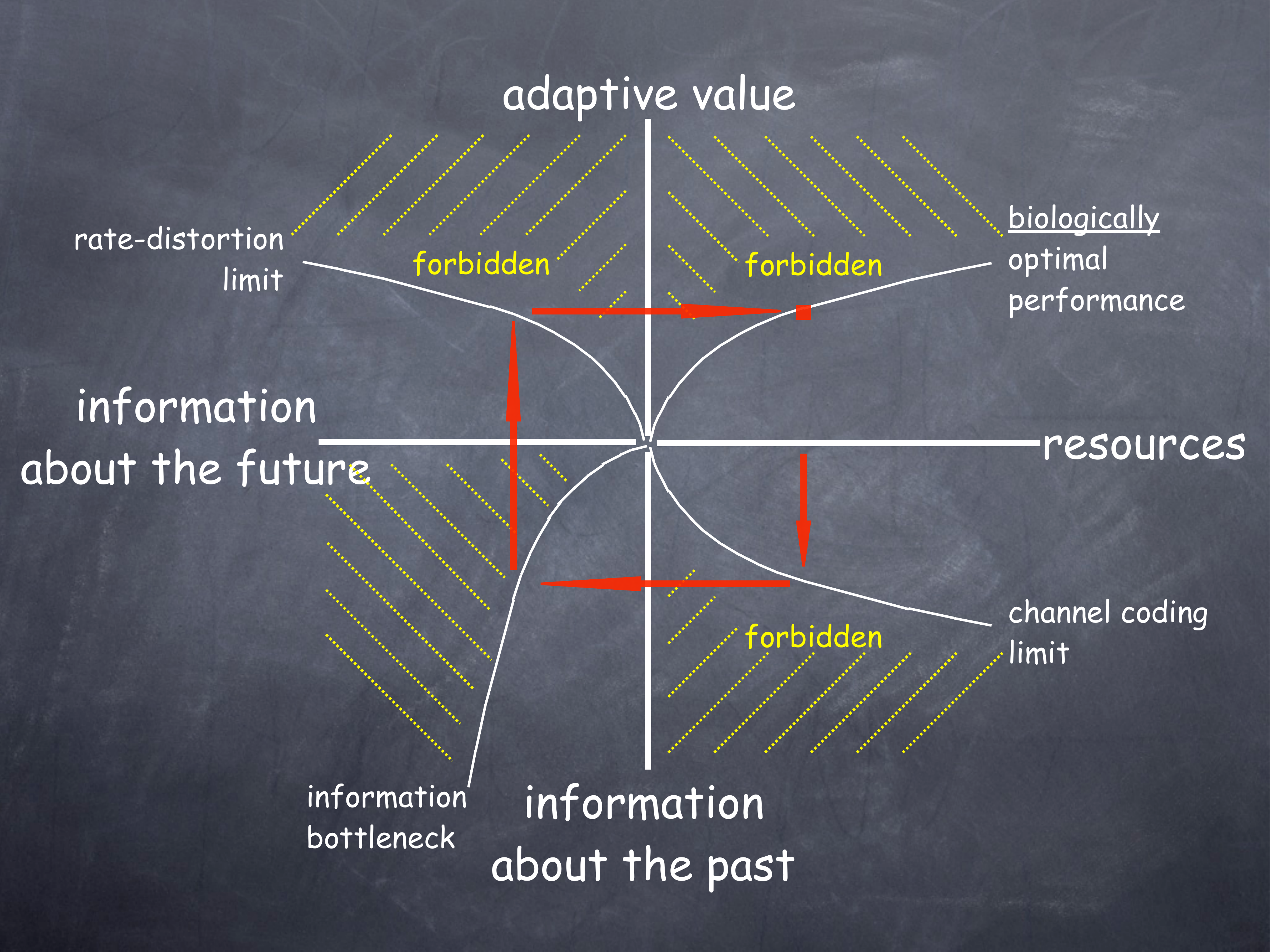}}
\caption{Connecting the different optimization principles.  Lines indicate curves of optimal performance, separating allowed from forbidden (hashed) regions of each quadrant.
In the upper right quadrant is the biological principle, maximizing fitness or adaptive value at fixed resources. But actions that achieve a given level of adaptive value require a minimum number of bits, and since actions occur after plans these are bits about the future (upper left).  On the other hand, the organism has to ``pay'' for bits, and hence there is a minimum resource costs for any representation of information (lower right).  Finally, given some bits (necessarily obtained from observations on the past), there is some maximum number of bits of predictive power (lower left).  To find a point on the biological optimum one can try to follow a path through the other three quadrants, as indicated by the arrows.}
\label{4quad}
\end{figure}

The fact that representations are built from data about the past but are useful only to the extent that they provide information about the future means that, for the organism,  the bits in the rate--distortion tradeoff are bits about the future, while the bits in the channel capacity tradeoff are bits about the past.  Thus the different tradeoffs we have been discussing---the biologically relevant trade between resources and adaptive value, the rate--distortion relation between adaptive value and bits, and the channel capacity trading of resources for bits---form three quadrants in a plane (Fig \ref{4quad}).  The fourth quadrant, which completes the picture, is a purely information theoretic tradeoff between bits about the past and bits about the future.

Assuming that the organism lives in a statistically stationary world, predictions ultimately are limited by the statistical structure of the data that the organism collects.  More concretely, if we observe a time series through a window of duration $T$ (that is, for times $-T < t \leq 0$), then to represent the data $X_{\rm past}$ we have collected requires $S(T)$ bits, where $S$ is the entropy, but the information that these data provide about the future $X_{\rm future}$ (i.e., at times $t > 0$) is given by some $I(X_{\rm past} ; X_{\rm future}) \equiv I_{\rm pred}(T) \ll S(T)$.  In particular, while for large $T$ the entropy $S(T)$ is expected to become extensive, the predictive information $I_{\rm pred}(T)$ always is subextensive \cite{BNT}.   Thus we expect that the data $X_{\rm past}$ can be compressed significantly into some internal representation $X_{\rm int}$ without losing too much of the relevant information about $X_{\rm future}$.  This problem---mapping $X_{\rm past} \rightarrow X_{\rm int}$ to minimize the information $I(X_{\rm int}; X_{\rm past})$ that we keep about the past while maintaining information $I(X_{\rm int}; X_{\rm future})$ about the future---is an example of the ``information bottleneck'' problem \cite{IB}.  Again there is a curve of optimal performance, separating the plane into allowed and forbidden regions.  Formally, we can construct this optimum by solving
\begin{equation}
\max_{X_{\rm past} \rightarrow X_{\rm int}}\left[I(X_{\rm int}; X_{\rm future}) - \lambda 
I(X_{\rm int}; X_{\rm past})\right],
\label{optprin}
\end{equation}
where $X_{\rm past} \rightarrow X_{\rm int}$ is the rule for creating the internal representation and $\lambda$ is a Lagrange multiplier.
 
We see that there are several different optimization principles, all connected, as schematized in Fig \ref{4quad}.  The biologically relevant principle is to maximum the fitness $F$ given some resource cost $C$.  But in order to take actions that achieve some mean fitness $F$ in a potentially fluctuating environment, the organism must have an internal representation $X_{\rm int}$ that provides some minimum amount of information $I(X_{\rm int}; X_{\rm future})$ about the future states of that environment; the curve $I(X_{\rm int}; X_{\rm future})$ vs. $F$ is a version of the rate--distortion curve.  Building and acting upon this internal representation, however, entails various costs, and these can all be assigned to the construction of the representation out of the (past) data as they are collected; the curve of $I(X_{\rm int}; X_{\rm past})$ vs. $C$ is an example of the channel capacity.  Finally, the information bottleneck principle tells us that there is an optimum choice of internal representation which maximizes $I(X_{\rm int}; X_{\rm future})$ at fixed $I(X_{\rm int}; X_{\rm past})$.   

The four interconnected optimization principles certainly have to be consistent with one another.  Thus, if an organism wants to achieve a certain mean fitness, it needs a minimum number of bits of predictive power, and this requires collecting a minimum number of bits about the past, which in turn necessitates some minimum cost. The possible combinations of cost and fitness---the accessible region of the biologically meaningful tradeoff in Fig \ref{bio_opt}---thus have a reflection in the ``information plane'' (the lower left quadrant of Fig \ref{4quad}) where we trade bits about the future against bits about the past.

The consistency of the different optimization principles means that the purely information theoretic tradeoff between bits about the future and bits about the past must constrain the biologically optimal tradeoff between resources and fitness.  We would like to make ``constrain'' more precise, and conjecture that under reasonable conditions organisms which operate at the biological optimum (that is, along the bounding curve in the upper right quadrant of Fig \ref{4quad}) also operate along the information theoretic optimum (the bounding curve in the lower left quadrant of Fig \ref{4quad}).  At the moment this is only a conjecture, but we hope that the relationships in Fig \ref{4quad} open a path to some more rigorous connections between information theoretic and biological quantities.

\section{A unifying principle?}

The optimization principle in Eq (\ref{optprin}) is very abstract; here we consider two concrete examples.  First imagine that we observe a Gaussian stochastic process [$x(t)$] that consists of a correlated signal [$s(t)$] in a background of white noise [$\eta(t)$].  
For simplicity, let's understand `correlated' to mean that $s(t)$ has an exponentially decaying correlation function with a correlation time $\tau_c$.
Thus, $x(t) = s(t) + \eta (t)$, where 
\begin{eqnarray}
\langle s(t) s(t') \rangle &=& \sigma^2 \exp\left(-{{|t-t'|}/{\tau_c}}\right)\\
\langle \eta (t) \eta (t')\rangle &=& {\cal N}_0\delta (t-t'),
\end{eqnarray}
and hence the power spectrum of $x(t)$ is given by
\begin{eqnarray}
\langle x(t) x(t')\rangle &\equiv& \int {{d\omega}\over{2\pi}} S_x(\omega) \exp\left[ -i\omega (t-t')\right]\\
S_x(\omega) &=& {{2\sigma^2\tau_c}\over{1+(\omega\tau_c )^2}} + {\cal N}_0 .
\end{eqnarray}
The full probability distribution for the function $x(t)$ is 
\begin{equation}
P[x(t)] = {1\over Z}
\exp\left[ -{1\over 2}\int dt\int dt'\, x(t) K(t-t') x(t')\right] ,
\label{gauss}
\end{equation}
where $Z$ is a normalization constant and the kernel
\begin{equation}
K(\tau ) = \int {{d\omega}\over{2\pi}} {1\over {S_x(\omega )}}\exp(-i\omega\tau ).
\end{equation}

If we sit at $t=0$, then $X_{\rm past} \equiv x(t<0)$ and $X_{\rm future}\equiv x(t>0)$.
In the exponential of Eq (\ref{gauss}), mixing between $X_{\rm past}$ and $X_{\rm future}$ is confined to a term which can be written as
\begin{equation}
\left[\int_{-\infty}^0 dt \, g(-t) x(t)\right] \times \left[ \int_0^\infty dt' \,g(t') x(t')\right] ,
\end{equation}
where $g(t) = \exp(-t/\tau_0)$, with $\tau_0 = \tau_c(1+\sigma^2\tau_c/{\cal N}_0)^{-1/2}$.
This means that the probability distribution of $X_{\rm future}$ given $X_{\rm past}$ depends only on $x(t)$ as seen through the linear filter $g(\tau )$, and hence only this filtered version of the past can contribute to $X_{\rm int}$ \cite{gaussIB}.

The filter $g(t)$ is exactly the filter that provides optimal separation between the signal $s(t)$ and the noise $\eta(t)$;
more precisely, given the data $X_{\rm past}$, if we ask for the best estimate of the signal $s(t)$, where ``best'' means minimizing the mean--square error, then this optimal estimate is just $y(t)$ 
\cite{wiener}.  Solving the problem of optimally representing the predictive information in this time series thus is identical to  the problem of optimally separating signal from noise.  

In contrast to these results for Gaussian time series with finite correlation times, consider what happens we look at a time series that has essentially infinitely long correlations.  Specifically, consider an ensemble of possible experiments in which points $x_{\rm n}$ are drawn independently and at random from the probability distribution $P(x|{\bf \vec\alpha})$, where $\bf\vec\alpha$ is a $K$--dimensional vector of parameters specifying the distribution.  At the start of each experiment these parameters are drawn from the distribution $P({\bf\vec\alpha})$ and then fixed for all $\rm n$.  Thus the joint distribution for many successive observations on $x$ on one experiment is given by
\begin{equation}
P(x_1, x_2 , \cdots , x_M) = \int d^K\alpha P({\bf\vec\alpha})\prod_{{\rm n}=1}^M
P(x_{\rm n}|{\bf \vec\alpha}).
\end{equation}
Now we can define $X_{\rm past} \equiv \{x_1, x_2 , \cdots , x_N\}$ and $X_{\rm future}\equiv \{x_{N+1}, x_{N+2} , \cdots , x_M\}$, and we can (optimistically) imagine an unbounded future, $M\rightarrow\infty$.  To find the optimal representation of predictive information in this case we need a bit more of the apparatus of the information bottleneck \cite{IB}.

It was shown in Ref \cite{IB} that an optimization problem of the form in Eq (\ref{optprin}) can be solved by probabilistic mappings $X_{\rm past} \rightarrow X_{\rm int}$ provided that the distribution which describes this mapping obeys a self--consistent equation,
\begin{equation}
P(X_{\rm int}|X_{\rm past}) = {1\over{Z(X_{\rm past}; \lambda)}}
\exp\left[ - {1\over \lambda} D_{\rm KL}(X_{\rm past}; X_{\rm int}) \right] ,
\label{ibeqn}
\end{equation}
where $Z$ is a normalization constant and $D_{\rm KL}(X_{\rm past}; X_{\rm int})$ is the Kullback--Leibler divergence between the distributions of $X_{\rm future}$ conditional on $X_{\rm past}$ and $X_{\rm int}$, respectively,
\begin{eqnarray}
 D_{\rm KL}(X_{\rm past}; X_{\rm int})
 &=& \int DX_{\rm future} P(X_{\rm future}|X_{\rm past}) 
 \nonumber\\
 &&\,\,\,\,\,\times\ln\left[
 {{P(X_{\rm future}|X_{\rm past})}\over{P(X_{\rm future}|X_{\rm int})}}\right] .
\end{eqnarray}
Since the future depends on our internal representation only because this internal representation is built from observations on the past, we can write
\begin{eqnarray}
P(X_{\rm future}|X_{\rm int}) &=& \int DX_{\rm past} P(X_{\rm future}|X_{\rm past}) \nonumber\\
&&\,\,\,\,\,\times P(X_{\rm past}|X_{\rm int})\\
P(X_{\rm past}|X_{\rm int}) &=& P(X_{\rm int}|X_{\rm past} ) {{P(X_{\rm past})}\over{P(X_{\rm int})}} ,
\end{eqnarray}
which shows that Eq (\ref{ibeqn}) really is a self--consistent equation for $P(X_{\rm int}|X_{\rm past})$.  To solve these equations it is helpful to realize that they involve integrals over many variables, since $X_{\rm past}$ is $N$ dimensional and $X_{\rm future}$ is $M$ dimensional.  In the limit that these numbers are very large and the temperature--like parameter $\lambda$ is very small, it is plausible that the relevant integrals are dominated by their saddle points.  

In the saddle point approximation one can find solutions for $P(X_{\rm int}|X_{\rm past})$ that have the following suggestive form.  The variable $X_{\rm int}$ can be thought of as a point in a $K$--dimensional space, and then the distributions $P(X_{\rm int}|X_{\rm past})$ are Gaussian, centered on locations ${\bf\vec\alpha}_{\rm est}(X_{\rm past})$, which are the maximum a posteriori Bayesian estimates of the parameters $\bf\vec\alpha$ given the observations $\{x_1 , x_2 , \cdots , x_N\}$.  The covariance of the Gaussian is proportional to the inverse Fisher information matrix, reflecting our certainty about $\bf\vec\alpha$ given the past data.  Thus, in this case, if we solve the problem of efficiently representing predictive information, then we have solved  the problem of learning the parameters of the probabilistic model that underlies the data we observe.

Signal processing and learning usually are seen as very different problems, especially from a biological point of view.  Building an optimal filter to separate signal from noise is a ``low--level'' task, presumably solved in the very first layers of sensory processing.  Learning is a higher level problem, especially if the model we are learning starts to describe objects far removed from raw sense data, and presumably happens in the cerebral cortex.  Returning to the problem of placing value on information, separating signal from noise by filtering and learning the parameters of a probabilistic model seem to be very different goals.  In the conventional biological view,  organisms carry out these tasks at different times and with different mechanisms because the kinds of information that one extracts in the two cases have different value to the organism.  What we have shown is that there is an alternative, and more unfiied, view.

There is a single principle---efficient representation of predictive information---that values all (predictive) bits equally but in some instances corresponds to filtering and in others to learning.  In this view, what determines whether we should filter or learn is not an arbitrary ``biological'' choice of goal or assignment of value, but rather the structure of the data stream to which we have access.  

\section{Neural coding of predictive information}

It would be attractive to have a direct test of these ideas.  We recall that neurons respond to sensory stimuli with sequences of identical action potentials or ``spikes,'' and hence the brain's internal representation of the world is constructed from these spikes \cite{spikes}.  More narrowly, if we record from a single neuron, 
then this internal representation $X_{\rm int}$ can be identified with a short segment of the spike train from that neuron, while $X_{\rm past}$ and $X_{\rm future}$ are the past and future sensory inputs, respectively.    The conventional analysis of neural responses focuses on the relationship between $X_{\rm past}$ and $X_{\rm int}$---trying to understand what features of the recent sensory stimuli are responsible for shaping the neural response.  In contrast, the framework proposed here suggests that we try to quantify the information $I(X_{\rm int}; X_{\rm future})$ that neural responses provide about the future sensory inputs \cite{more_general}.  More specifically,  to test the hypothesis that the brain generates maximally efficient representations of predictive information, we need to measure directly both $I(X_{\rm int}; X_{\rm future})$ and $I(X_{\rm int}; X_{\rm past})$, and see whether in a given sensory environment the neural representation $X_{\rm int}$ lies near the optimal curve predicted from Eq (\ref{optprin}).

 \begin{figure*}
 \centerline{\includegraphics[width=1.5\columnwidth]{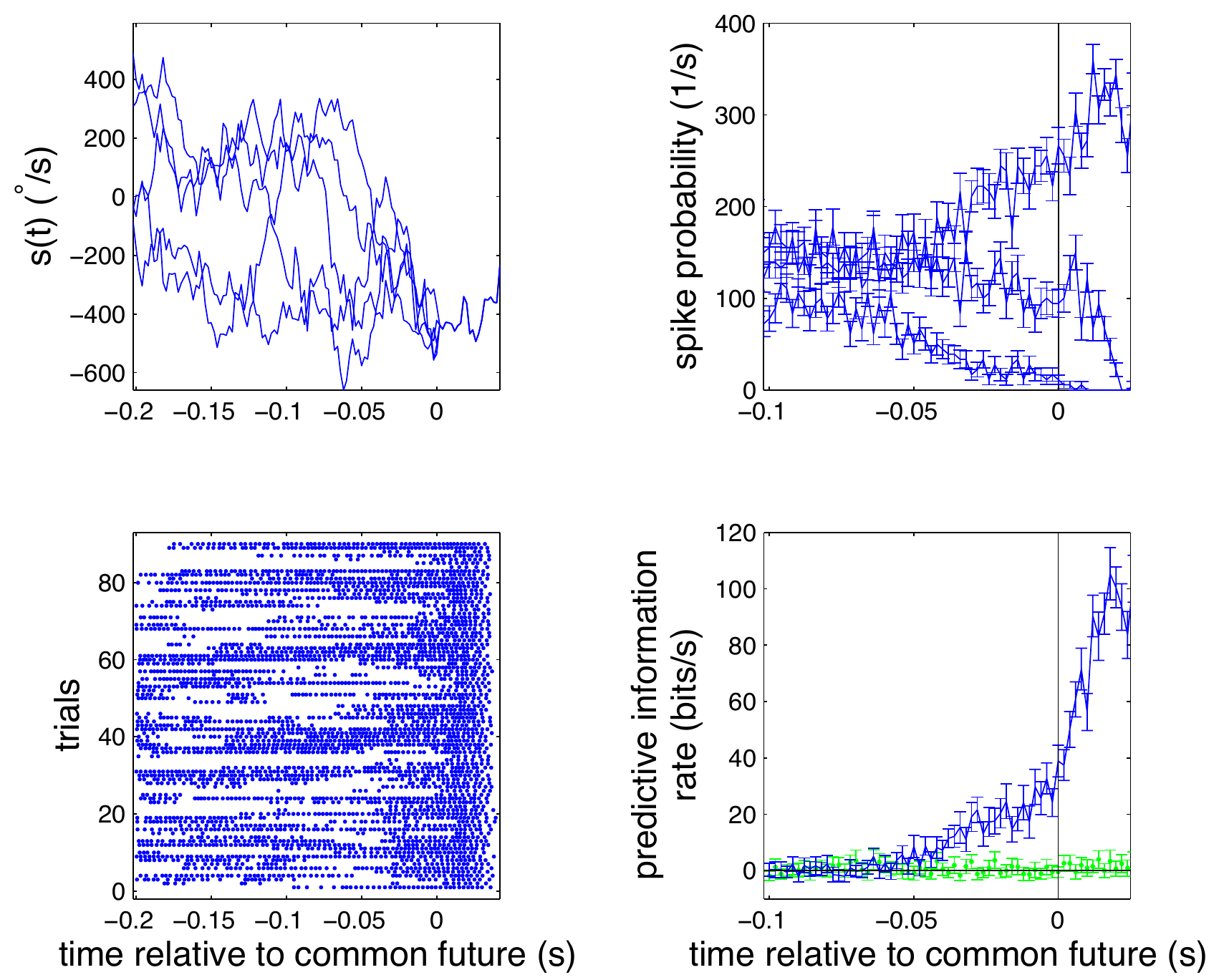}}
 \caption{Trajectories with a common future and their neural representation.
 Top left: Sample trajectories $s_{\rm k} (t)$  designed to converge on a common future at $t=0$, as explained in the text.  Units are angular velocity, since these signals will be used to drive the motion--sensitive visual neuron H1.  The correlation time  $\tau_c = 0.05\,{\rm s}$ and the variance $\langle s^2 \rangle = (200^\circ/{\rm s})^2$.
 Bottom left:  Responses of the blowfly H1 neuron to ninety different trajectories of angular velocity vs. time $s_{\rm k}(t)$ converge on a common future at $t=0$; each dot represents a single spike generated by H1 in response to these individual signals. Stimulus delivery and recordings as described in Ref \cite{lewen+al_01}.
 Top right:  Probability per unit time of observing a spike in response to trajectories that converge on three different common futures.  At times long before the convergence, all responses are drawn from the same distribution and hence have same spike probability within errors.  Divergences among responses begin at a time $\sim \tau_c$ prior to common future.  This divergence means that the neural responses carry information about the particular common future, as explained in the text. Error bars are standard errors of the mean, estimated by bootstrapping.
 Bottom right: Information in single time bins about the identity of the future, normalized as an information rate.  Blue points are from the real data, and green points are from shuffled data that should have zero information.}
 \label{bigfig}
\end{figure*}

It would seem that to measure $I(X_{\rm int}; X_{\rm future})$ we would have to understand the structure of the code by which spike trains represent the future; the same problem arises even with $I(X_{\rm int}; X_{\rm past})$.  
In fact there is a more direct strategy \cite{strong+al_98}.
The essential idea behind direct measurements of neural information transmission \cite{strong+al_98} is to use the (ir)reproducibility of the neural response to repeated presentations of the same dynamic sensory signal.  If we think of the sensory stimulus as a movie that runs from time $t=0$ to $t=T$, we can run the movie repeatedly  in a continuous loop.  Then at each moment $t$ we can look at the response $R\equiv X_{\rm int}$ of the neuron, and if there are enough repetitions of the movie we can estimate the conditional distribution $P(X_{\rm int}|t)$; the entropy $S_n(t)$ of this distribution measures the ``noise'' in the neural response.  On the other hand, if we average over the time $t$ we can estimate $P(X_{\rm int})$, and the entropy $S_{\rm total}$ of this distribution measures the capacity of the neural responses to convey information.  In the limit of large $T$ ergodicity allows us to identify averages over time with averages over the distribution out of which the stimulus movies are being drawn, and then 
\begin{equation}
I  = S_{\rm total}-{1\over T} \int_0^T S_n(t)
\end{equation}
 is the mutual information between sensory inputs and neural responses.  Since the neuron responds causally to its sensory inputs, the information that it carries about these inputs necessarily is information about the past, $I(X_{\rm int}; X_{\rm past})$ \cite{no_feedback}. Note that this computation does not require us to understand how to read out the encoded information, or even to know which features of the sensory inputs are encoded by the brain.
 
More careful analysis makes clear that the strategy in Ref \cite{strong+al_98} measures the information which $X_{\rm int}$ provides about whatever aspects of the sensory stimulus are being repeated.  For example,  if we have a movie with sound and we repeat the video but randomize the audio, then following the analysis of Ref \cite{strong+al_98} we would measure the information that neurons carry about their visual and not auditory inputs.  Thus to measure $I(X_{\rm int}; X_{\rm future})$ we need to generate sensory stimuli that are all drawn independently from the same distribution but are constrained to lead to the same future, and then reproducible neural responses to these stimuli will reflect information about the future.  
This can be done by a variety of methods.
  
Consider a time dependent signal $s_{\rm k}(t)$ generated on repeat $\rm k$ as
\begin{equation}
\tau_c{{ds_{\rm k}(t)}\over{dt}} + s_{\rm k}(t) = \xi_{\rm k}(t) ,
\label{langevin}
\end{equation}
where $\xi_{\rm k}(t<0) = \xi_0(t)$ for all $\rm k$, while each $\xi_{\rm k}(t>0)$ is drawn independently; in the simplest case $\xi(t)$ has no correlations in time (white noise).  Then the correlation function of the signal becomes
\begin{equation}
\langle s_{\rm k}(t) s_{\rm k}(t')\rangle \propto\exp(-|t-t'|/\tau_c),
\end{equation}
and all $s_{\rm k}(t<0)$ are identical.  Now take the trajectories and reverse the direction of time.   The result is an ensemble of trajectories that lead to the same future but are otherwise statistically independent, as in upper left panel in Fig \ref{bigfig}.

We have used the strategy outlined above to explore the coding of predictive information in the fly visual system, returning to the neuron H1.   The extreme stability of recordings from H1 
has been exploited in experiments where we deliver motion stimuli by physically rotating the fly outdoors in a natural environment rather than showing movies to a fixed fly \cite{lewen+al_01}, and this is the path that we follow here.

We have generated angular velocity trajectories $s(t)$ with a variance $\langle s^2\rangle = (200\,^\circ/{\rm s})^2$ and a correlation time $\tau_c = 0.05\,{\rm s}$ by numerical solution of Eq (\ref{langevin}).  We choose nine such segments at random to be the common futures, and then follow the construction leading to the upper left panel in Fig \ref{bigfig} to generate ninety independent trajectories for each of these common futures.  These trajectories are used as angular velocity signals to drive rotation of a blowfly {\em Calliphora vicina} mounted on a motor drive as in Ref \cite{lewen+al_01} while we record the spikes generated by the H1 neuron.

The lower left panel in Figure \ref{bigfig} shows examples of the spike trains generated by H1 in response to independent stimuli that converge on a common future.    Long before the convergence, stimuli are completely different on every trial, and hence the neural responses are highly variable.  As we approach the convergence time, stimuli on different trials start to share features which are predictive of the common future, and hence the neural responses become more reproducible.  Importantly, stimulus trajectories that converge on different common futures generate responses that are not only reproducible but also distinct from one another, as seen in the upper right of Fig \ref{bigfig}.  Our task now is to quantify this distinguishability by estimating $I(X_{\rm int}; X_{\rm future})$.

Imagine dividing time into small bins of duration $\Delta\tau$.  For small $\Delta\tau$, we observe either one spike or no spikes, so the neural response is  a binary variable.  If we sit at one moment in time relative to the convergence on a common future, we observe this binary variable, and it is associated with one of the nine possible futures.  Thus there is a $2\times 9$ table of futures and responses, and it is straightforward to use the experiments to fill in the frequencies with which each of these response/future combinations occurs. With $810$ samples to fill in the frequencies of $18$ possible events, we have reasonably good sampling and can make reliable estimates of the mutual information, with error bars, following the methods of Refs \cite{strong+al_98,slonim+al_05}.  In each time bin, then, we can estimate the information that the neural response provides about the future, and we can normalize this by the duration of the bin $\Delta\tau$ to obtain an information rate, as shown at the lower right in Fig \ref{bigfig}.  We see that information about sensory signals in the future ($t>0$) is negligible in the distant past, as it must be, with the scale of the decay set by the correlation time $\tau_c$.  The local information rate builds up as we approach $t=0$, peaking for this stimulus ensemble at $\sim 30\,{\rm bits/s}$.

\begin{figure}[tb]
\vskip 0.2 in
\centerline{\includegraphics[width=0.9\columnwidth]{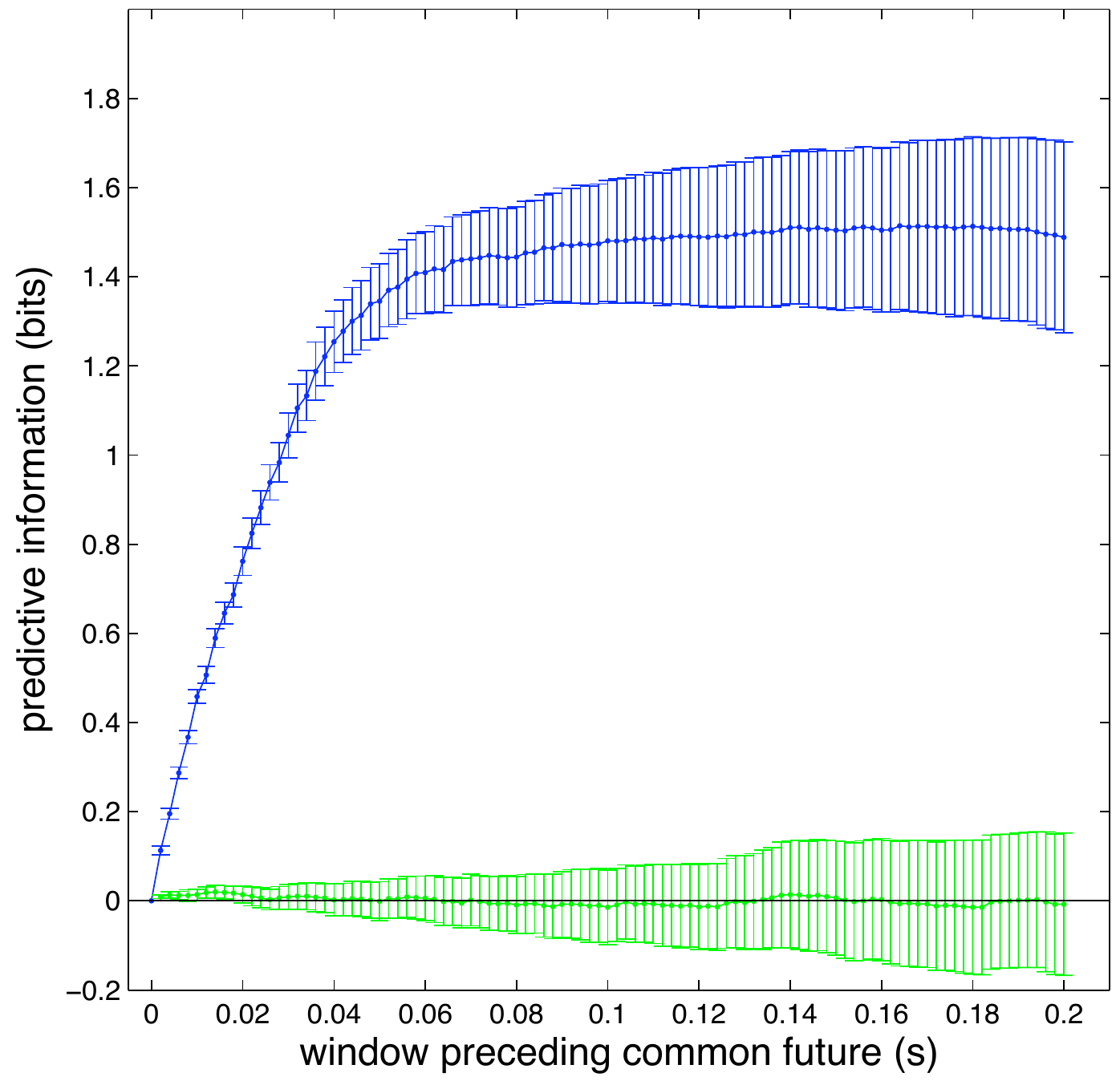}}
\caption{Information about the future carried by spikes in a window that ends at the time of convergence onto a common future.  Essentially this is the integral of the information rate shown in at the lower right in Fig \ref{bigfig}, but error bars must be evaluated carefully.  Shown in green are the results for shuffled data, which should be zero within errors if our estimates are reliable.}
\label{summedinfo}
\end{figure}

The results in Fig \ref{bigfig} provide a moment by moment view of the predictive information in the neural response, but we would like a slightly more integrated view:  if we sit at $t=0$ and look back across a window of duration $T$, how much predictive information can we extract from the responses in this window?  The complication in answering this question is that the neural response across this window is a $T/\Delta\tau$--letter binary word, and the space of these words is difficult to sample for large $T$.  The problem could be enormously simpler, however, if the information carried by each spike were independent, since then the total information would be the integral of the local information rate.  This independence isn't exactly true, but it isn't  a bad approximation under some conditions \cite{brenner+al_00b}, and we adopt it here.  Then the only remaining technical problem is to be sure that small systematic errors in the estimate of the local information rate don't accumulate as we compute the time integral, but this can be checked by shuffling the data and making sure that the shuffled data yields zero information.  The results of this computation are shown in Fig \ref{summedinfo}.

Should we be surprised by the fact that the neural response from this one single neuron carries somewhat more than one bit of information about the future?  Perhaps not.  This is, after all, a direction selective motion sensitive neuron, and because of the correlations the direction of motion tends to persist; maybe all we have found is that the neuron encodes the current sign of the velocity, and this is a good predictor of the future sign, hence one bit, with perhaps a little more coming along with some knowledge of the speed.  We'd like to suggest that things are more subtle.

Under the conditions of these experiments, the signal--to--noise in the fly's retina is quite high.  As explained in Ref \cite{fairhall+al_01}, we can think of the fly's eye as providing an essentially perfect view of the visual world that is updated with a time resolution $\Delta t$ on the scale of milliseconds.  But if we are observing a Gaussian stochastic process with a correlation time of $\tau_c$, then the limit to prediction is set by the need to extrapolate across the `gap' of duration $\Delta t$.  Since the exponentially decaying correlation function corresponds to a Markov process, this limiting predictive information is calculable simply as the mutual information between two samples separated by the gap; the result is
\begin{equation}
I_{\rm pred}= {1\over 2}\log_2\left[{1\over{ 1 - \exp(-2\Delta t /\tau_c)}}\right] .
\end{equation}
Plugging in the numbers, we find that, for these stimuli, capturing roughly one bit of predictive information depends in an essential way on the system have a time resolution of better than ten milliseconds, and the observed  predictive information requires resolution in the $3-4\,{\rm ms}$ range.

When we look back at a window of the neural response with duration $T$, we expect to gain $R_{\rm info}T$ bits of information about the stimulus \cite{careful}, and as noted above this necessarily is information about the past.  Thus the {\cal x}--axis of Fig \ref{summedinfo}, which measures the duration of the window, can be rescaled, so that the whole figure is a plot of information about the future vs information about the past, as in the lower left quadrant of Fig \ref{4quad}.  Thus we can compare this measure of neural performance with the information bottleneck limit derived from the statistical structure of the visual stimulus itself; since the stimulus is a Gaussian stochastic process, this is straightforward \cite{chechik+al_05}, and as above we assume that the fly has access to a perfect representation of the velocity vs. time, updated at multiples of the time resolution $\Delta t$.  With $\Delta t$ in the range of $3-4\,{\rm ms}$ to be consistent with the total amount of predictive information captured by the neural response, we find that the performance of the neuron always is within a factor of two of the bottleneck limit, across the whole range of window sizes.

\section{Discussion}

We should begin our discussion by reminding the reader that, more than most papers, this is a report on work in progress, intended to capture the current state of our understanding rather than to draw firm conclusions.

{\em Efficient representation?}  Although one should be careful of glib summaries, it does seem that the fly's visual system offers concrete evidence of the brain building representations of the sensory world that are efficient in the sense defined by information theory.  The absolute information rates are large (especially in comparison to prior expectations in the field!), and there are many signs that the coding strategy used by the brain is matched quantitatively to the statistical structure of sensory inputs, even as these change in time.  This matching, which gives us a much broader view of  ``adaptation'' in sensory processing, has now been observed directly in many different systems \cite{wark+al_07}.

{\em Why are these bits different from all other bits?} Contrary to widespread views in the neuroscience community, information theory does give us a language for distinguishing relevant from irrelevant information.  We have tried to argue that, for living organisms, the crucial distinction is predictive power.  Certainly data without predictive power is useless, and thus `purifying' predictive from non--predictive bits is an essential task.  Our suggestion is that this purification may be more than just a first step, and that providing a maximally efficient representation of the predictive information can be mapped to more biologically grounded notions of optimal performance.  Whether or not this general argument can be made rigorous, certainly it is true that extracting predictive information serves to unify the discussion of problems as diverse as signal processing and learning.

{\em  A new look at the neural code?}  The traditional approach to the the analysis of neural coding tries to correlate (sometimes in the literal mathematical sense) the spikes generate by neurons with particular features of the sensory stimulus.  But, because the system is causal, these features must be features of the organism's recent past experience.  Our discussion of predictive information suggests a very different view, in which we ask how the neural response represents the organism's future sensory experience.  Although there are many things to be done in this direction, we find it exciting that one can make rather direct measurements of the predictive power encoded in the neural response.

From an experimental point of view, the most compelling success would be to map neural responses to points in the information plane---information about the future vs information about the past---and find that these points are close to the theoretical optimum determined by the statistics of the sensory inputs and the information bottleneck.  We are close to being able to do this, but there is enough uncertainty in our estimates of information (recall that we work only in the approximation where spikes carry independent information) that we are reluctant to put theory and experiment on the same graph.  Our preliminary result, however, is that theory and experiment agree within a factor of two, encouraging us to look more carefully.

\acknowledgments{We thank N Brenner, G Chechik, AL Fairhall, A Globerson, R Koberle, GD Lewen, I Nemenman, FC Pereira, E Schneidman, SP Strong and Y Weiss for their contributions to  work reviewed here, and T Berger for the invitation to speak at ISIT06.   This work was supported in part by the National Science Foundation, though Grants IIS--0423039 and PHY--0650617, and by the Swartz Foundation.}


\begin{thebibliography}{99}
\bibitem{gaussIB}
The information bottleneck in the case of Gaussian distributions admits a more general discussion, in particular identifying those instances in which $X_{\rm int}$ depends only on a limited set of linear projections of the input variable ($X_{\rm past}$ in the present discussion).  See Ref \cite{chechik+al_05} for details.
\bibitem{more_general}
There is a literature which focuses on the predictive nature of neural responses, but only in very specific situations such as the prediction of rewards.  The arguments presented above suggest that prediction is a much more general problem.
\bibitem{no_feedback}
This step depends on the design of the experiment.  In fully natural conditions, the outputs of sensory neurons serve to guide motor behavior, and hence the future sensory inputs depend explicitly on the internal representation.  Analysis of this closed loop situation is beyond the scope of our theoretical discussion above, so we consider here experiments in which sensory stimuli are independent of the neural output.
\bibitem{careful}
Strictly speaking this is correct only for large $T$, but in the approximation that each spike provides independent information this proportionality is exact.
\end{thebibliography}
\end{document}